\documentclass{aa}
\usepackage{graphicx}
\usepackage{txfonts}
\usepackage{natbib}
\bibpunct{(}{)}{;}{a}{}{,}
\usepackage{longtable}
\usepackage{lscape}
\usepackage{times}
\usepackage{textcomp}

\begin{document}

\title{Long, depolarising H$\alpha$-filament towards the Monogem ring\thanks{The FITS images of the radio maps are available in electronic form at the CDS via anonymous ftp to ...}}
%
\author{Wolfgang~Reich\inst{1}, Patricia~Reich\inst{1} and Xiaohui~Sun\inst{2} }
\titlerunning{G203.7+11.5}
\authorrunning{W. Reich et al.}

\offprints{wreich@mpifr.de}

\institute{
  Max-Planck-Institut f\"{u}r Radioastronomie, Auf dem H\"{u}gel 69,
  53121 Bonn, Germany; \\{\it wreich@mpifr-bonn,mpg.de, preich@mpifr-bonn.mpg.de} 
  \and Department of Astronomy, Yunnan University, and Key Laboratory of Astroparticle 
  Physics of Yunnan Province, Kunming 650091, China; {\it xhsun@ynu.edu.cn} }

\date{Received; accepted}

\abstract
{In soft X-rays, the Monogem ring is an object with a diameter of $25\degr$ located in the Galactic anti-centre. 
It is believed to be a faint, evolved, local supernova remnant. 
The ring is also visible in the far-ultraviolet, and a few optical filaments are
related. It is not seen at radio wavelengths, as other large supernova remnants are. 
} 
{We study a narrow about $4\fdg5$ long, faint H$\alpha$-filament, G203.7+11.5,
that is seen towards the centre of the Monogem ring. It causes depolarisation and excessive Faraday 
rotation of radio polarisation data.   
}
{Polarisation observations at $\lambda$11\ cm and  $\lambda$21\ cm with the Effelsberg 
100-m telescope were analysed 
in addition to $WMAP$ data, extragalactic rotation measures, and H$\alpha$ data. A Faraday-screen 
model was applied.}
{From the analysis of the depolarisation properties of the H$\alpha$  filament, we derived a line-of-sight 
magnetic field, $B_{||}$, of 26$\pm5\mu$G  for a distance
of 300~pc and an electron density, $n_\mathrm{e}$, of 1.6~cm$^{-3}$. The absolute largest rotation measure of 
G203.7+11.5 is  -86$\pm3$~rad~ m$^{-2}$, where the magnetic field direction has the opposite sign
from the large-scale Galactic field.
We estimated the average synchrotron emissivity at  $\lambda$21\ cm up to 300~pc distance towards 
G203.7+11.5 to about 1.1~K $T_\mathrm{b}$/kpc, which is higher than typical Milky Way values.}
{The magnetic field within G203.7+11.5 is unexpected in direction and strength. Most likely, the filament is
related to the Monogem-ring shock, where interactions with ambient clouds may cause local magnetic field 
reversals. We confirm earlier findings of an enhanced but direction-dependent local synchrotron emissivity. }


\maketitle
\section{Introduction}

X-rays from the Monogem ring were first detected during a rocket flight in 1969 \citep {Bunner70}.
It is seen as a bright soft X-ray object with a diameter of $25\degr$ in the {\it ROSAT} All-sky survey \citep{Voges99}, 
where its centre is at $l,b \sim 201\degr, +12\degr$.  The Monogem ring was discussed
in some detail by \citet{Plucinsky96}, who concluded that it is a large evolved supernova remnant 
(SNR) in the adiabatic phase of evolution that expands in a very low-density interstellar medium (ISM). 
More recent X-ray observations of the Monogem ring with {\it SUZAKU} led to a refined
analysis of the SNR parameters \citep{Knies18}. 

Far-ultraviolet (FUV) emission was  detected by \citet{Kim07}, where C IV and other lines indicate
interaction details of the Monogem ring with the ambient ISM. Two optical filaments associated with
the Monogem ring have been identified by \citet{Reimers84} and by \citet{Weinberger06}. 
The spectra of both filaments  show excitation by a slow shock, as expected from an evolved SNR. 
\citet{Thorsett03} showed that the pulsar PSR B0656+14 is located close to the centre of the 
Monogem ring, and the authors concluded that both objects result from a supernova explosion that occurred about one hundred 
thousand years ago at a distance of about 300~pc. 

The Monogem ring shows no radio emission, as expected for older SNRs in the metre or 
decimetre range, where synchrotron emission is the dominating process.  This was explained by  
\citet{Plucinsky96} by the unusually low ISM density. However, \citet{Vallee93} pointed out
that the Monogem ring and similar large local features may significantly contribute to the observed
rotation measure ($RM$) of extragalactic sources, which are the basis for magnetic field models of the
Milky Way.    

\begin{figure}
\centering
\includegraphics[angle=0, width=0.49\textwidth]{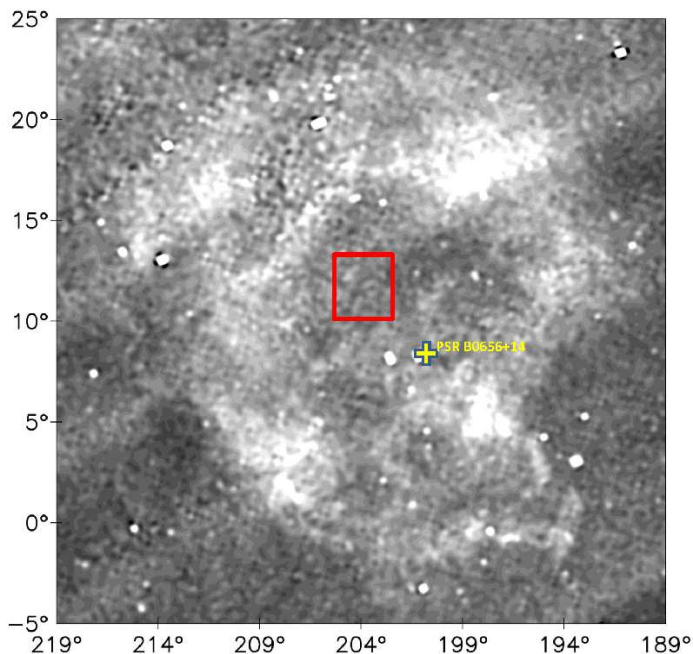}
\caption{Overview of the Monogem ring in soft-X-rays as observed by ROSAT (0.1- 0.4~keV band, \citep{Voges99}) with the related PSR B0656+14. The observed radio field is indicated.} 
\label{ROSAT}
\end{figure}

During the observations for the $\lambda$21\ cm Effelsberg Medium Latitude Survey (EMLS)
\citep{Uyaniker98, Uyaniker99, Reich04}, we found  numerous  
depolarised canal-like structures in the polarisation maps. For some of them,  we made follow-up observations at  $\lambda$11\ cm 
to study the origin of the canals. A prominent long canal is seen close to the direction of the centre 
of the Monogem ring. The filament has a clear counterpart in the all-sky H$\alpha$ map compiled by \citet{Finkbeiner03}.
Its properties are discussed in the following. We show ROSAT data of the Monogem ring in Fig.~\ref{ROSAT},
where we indicate the position of PSR B0656+14 and the area we observed at radio frequencies.

In Sect.~2 we describe the radio and H$\alpha$ data that we used and the zero-level calibration of the $\lambda$11\ cm polarisation
data. Section~3 presents the maps of our observations. A Faraday-screen analysis is given in Sect.~4. We discuss
the physical 
properties of G203.7+11.5 in Sect.~5, where we also derive the local synchrotron emissivity in the direction of the Monogem ring.  
Section~6 gives a summary.

\section{Data}

\subsection{Effelsberg  $\lambda$21\ cm data}

\begin{figure}
\centering
\includegraphics[angle=-90, width=0.49\textwidth]{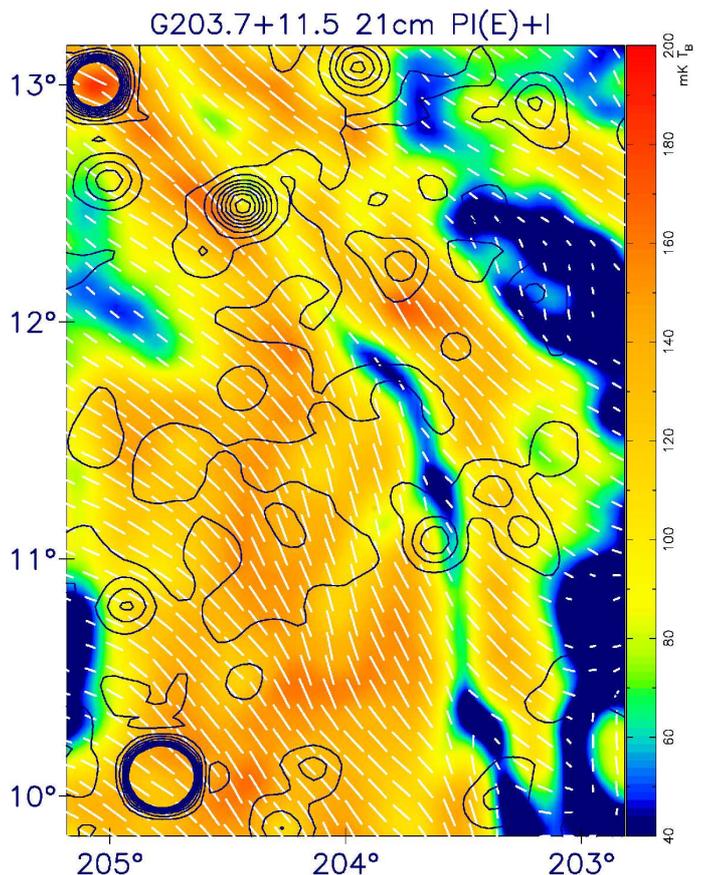}
\caption{Colour-coded $\lambda$21\ cm polarised emission (on an absolute zero-level)
  observed with the Effelsberg 100-m radio telescope. The angular resolution is $9\farcm4$. Overlaid contours show $\lambda$21\ cm
absolute total intensities starting at 3.8~K $T_\mathrm{b}$
  in steps of 100~mK $T_\mathrm{b}$. The bars are in E-field direction, and their length is proportional to the polarised intensity. 
} 
\label{21cm}
\end{figure}

G203.7+11.5 is visible on the EMLS $\lambda$21\ cm maps as a long, narrow 
polarisation depression (canal) that extends  from {\it{l,b}} $\sim 204\degr, +12\degr$ to {\it{l,b}} $\sim 203\fdg3, +10\degr$. 
We used total-intensity and linear-polarisation data
from an unpublished section of the EMLS. When completed, the EMLS will cover the northern Galactic 
plane $\pm20\degr$ in latitude at an angular resolution of $9\farcm4$. Table~\ref{ObsTab} lists some technical details. 
Missing large-scale emission components and the absolute zero-levels of the EMLS were provided by 
the Stockert 25-m telescope $\lambda$21\ cm total-intensity survey of the northern sky  
\citep{Reich82, Reich86} and the DRAO 26-m linear-polarisation survey \citep{Wolleben06}. 
Adding large-scale structures from these surveys provides an absolute zero-level for total intensities and 
linear polarisation. The $\lambda$21\ cm polarised-intensity (${PI}$) map with overlaid total-intensity contours
is shown in Fig.~\ref{21cm}. $PI$ is calculated from the observed Stokes parameter $U$ and $Q$ as
$PI = \sqrt{U^2+Q^2-1.2\sigma_{U,Q}^{2}}$, including zero-bias correction. $\sigma$ is listed in 
Table~\ref{ObsTab}. 

\begin{table*}[thp]
\caption{Observational parameters}

\label{tab1}
\vspace{-1mm}
\centering
\begin{tabular}{lrrr}
\hline\hline
\multicolumn{1}{c}{} &\multicolumn{1}{c}{}  & \multicolumn{1}{c}{}\\
Data                           &Effelsberg $\lambda$11\ cm   &EMLS $\lambda$21\ cm  \\
\hline             
Frequency [MHz]                                  &2639                         &1408 \\
Bandwidth [MHz]                                 &80                           &20   \\
HPBW[$\arcmin$]                                    &4.3                          &9.4 \\
Main Calibrator                                      &3C286                        &3C286 \\
Flux Density of 3C286 [Jy]                                 &11.5                         &14.4 \\
Polarisation Percentage of 3C286 [\%]                       &9.9                       &9.3\\
Polarisation Angle of 3C286 [$\degr$]                        &33                           &32 \\
Number of Coverages                         & $<$16                       &min. 1(B)+1(L) \\
Integration Time [s]                               & $<$16                            & $\geq$ 2 \\
r.m.s. ($I/U,Q$)[mK $T_\mathrm{b}$]                &4/2                             &15/8 \\
\hline\hline
\end{tabular}
\label{ObsTab}
\vspace{-1mm}
\end{table*}

\subsection{New Effelsberg $\lambda$11\ cm observations}

Radio continuum and linear polarisation observations of G203.7+11.5 were made at 
$\lambda$11\ cm with the Effelsberg 100-m radio telescope. The observations started in 
1997/1999 and were completed with an improved $\lambda$11\ cm receiver in 2007. The layout of 
the Effelsberg  $\lambda$11\ cm receiving system has been described by \citet{Uyaniker04}.  
The receiving system available in 1997/1999 was upgraded in 2005 with new low-noise amplifiers and an 
eight-channel IF-polarimeter. Its channels are 10~MHz wide, and a  broad-band channel provides 
80~MHz bandwidth in addition, which we used when no radio interference (RFI) was visible 
in the observing band.

\begin{figure}
\centering
\includegraphics[angle=-90, width=0.49\textwidth]{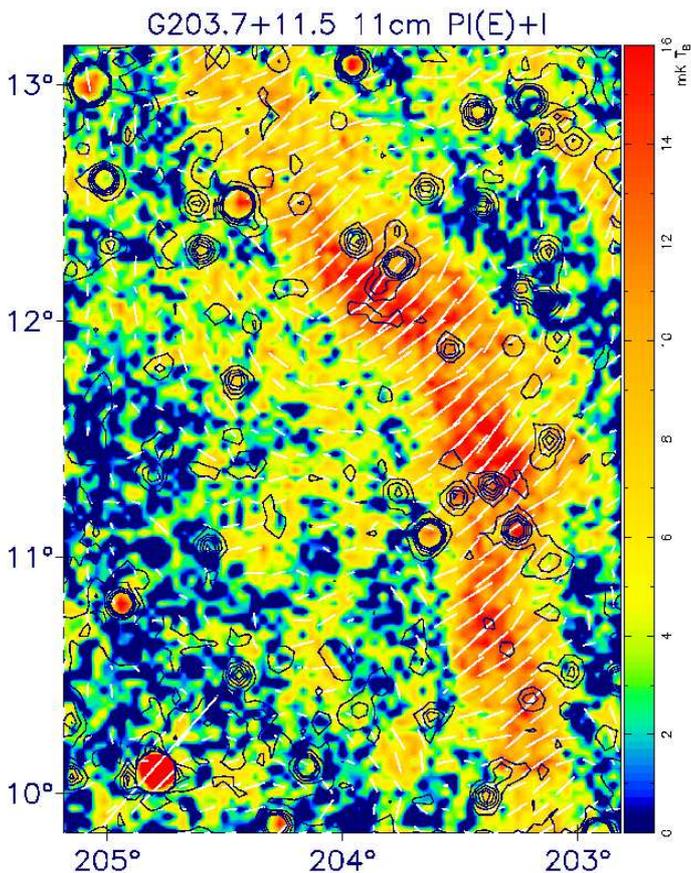}
\caption{As Fig.~\ref{21cm}, but displaying $\lambda$11\ cm polarised emission as
  observed with the Effelsberg 100-m radio telescope without large-scale emission. The angular resolution is $4\farcm3$. 
  Overlaid are 
  $\lambda$11\ cm total-intensity contours starting at 10~mK $T_\mathrm{b}$  in steps of 20~mK
  $T_\mathrm{b}$. 
  }
\label{11cm}
\end{figure}

A field of $2\fdg33 \times 3\fdg33$ was mapped by raster scans in
Galactic longitude and latitude direction.  Altogether, eight maps in each direction
were observed. However, about 50\% of the observations could not or only partly be used 
because of RFI, solar side-lobe distortions, or bad weather. Some observational parameters are listed 
in Table~\ref{ObsTab}. The raw data were reduced and calibrated by standard NOD2-based 
methods used for continuum and polarisation observations with the Effelsberg 100-m telescope 
(e.g. \citet{Reich9011}). We present the Effelsberg map of the $\lambda$11\ cm observations 
in Fig.~\ref{11cm}.   

\begin{figure}
\centering
\includegraphics[angle=-90, width=0.49\textwidth]{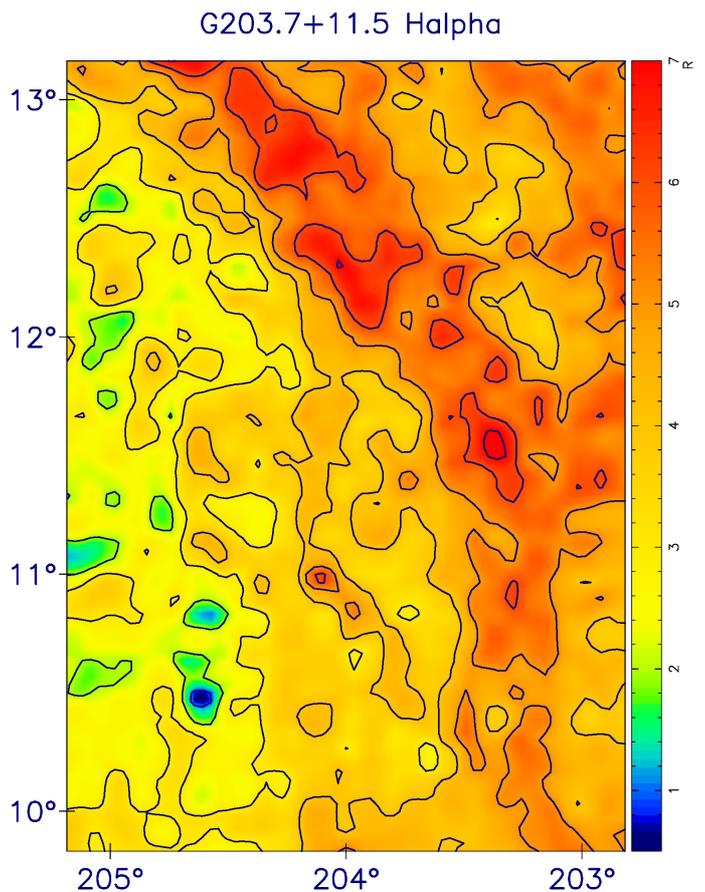}
\caption{H$\alpha$ emission extracted from the all-sky map of \citet{Finkbeiner03}.
Overlaid  H$\alpha$ contours extend from 1~Rayleigh to 7~Rayleigh in steps of 1~Rayleigh. 
} 
\label{Ha}
\end{figure}

\begin{figure}
\centering
\includegraphics[angle=-90, width=0.49\textwidth]{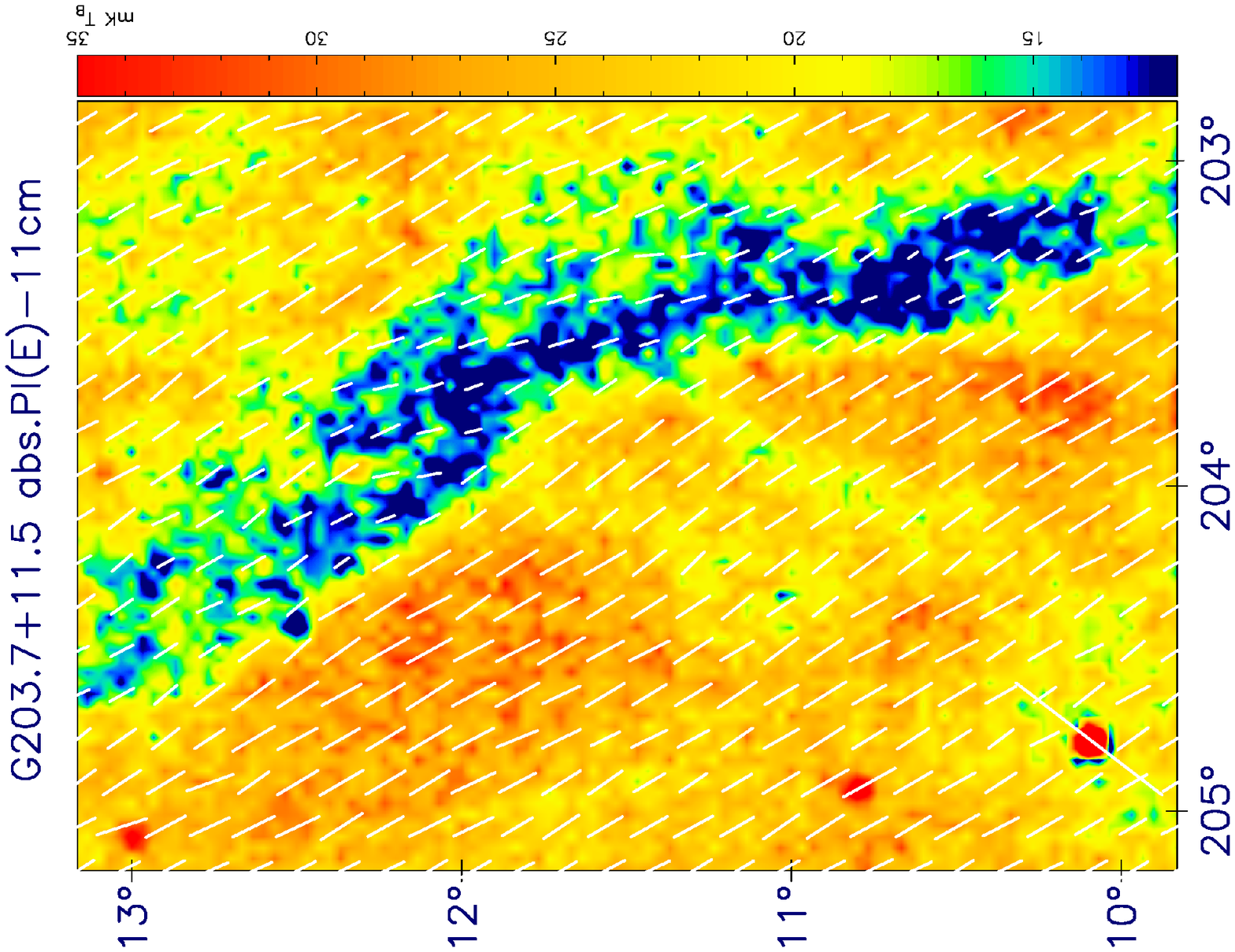}
\caption{Colour-coded Effelsberg $\lambda$11\ cm polarised emission as in Fig.~\ref{11cm} at an absolute 
zero-level, which was extrapolated from $WMAP$ data (see Sect. 2.4).
  }
\label{11cm+K9}
\end{figure}

\subsection{H$\alpha$ emission}

There is a clear H$\alpha$ counterpart of the polarised $\lambda$11\ cm emission 
in the all-sky H$\alpha$ map with 6$\arcmin$ angular resolution combined by \citet{Finkbeiner03}, which is a combination 
of various H$\alpha$ surveys. The  H$\alpha$ map (Fig.~\ref{Ha}) shows that the faint filament extends from {\it l,b} $\sim 
204\fdg5, +13\degr$ to $\sim 203\fdg1, +10\degr$  with intensity maxima around 
6 to 7~Rayleigh that become slightly fainter at its ends.  The intensity gradient is moderate 
and increases towards the upper right area of the map. The H$\alpha$ filament exceeds the diffuse large-scale emission by  about 3 to 4~Rayleigh 
and has a width of about $30\arcmin$ (HPBW).

\subsection{Absolute zero-level of the $\lambda$11\ cm polarisation data}

Interferometric data miss short spacings, and single-dish maps are usually set to an arbitrary zero-level. 
As discussed by \citet{Reich06} and others, polarisation data without restored large-scale emission make 
any interpretation of polarisation features unreliable when they result from radiation transfer and
not from emitting sources. Missing large-scale structures of polarisation maps from the magnetised 
interstellar medium will otherwise cause misinterpretations.

The observed  $\lambda$11\ cm Effelsberg maps of Stokes parameters $U$ and $Q$ were set to zero at their boundaries and thus miss
polarised emission from components exceeding about $2\degr$ to $3\degr$ in extent. 
The zero-level problem of single-dish telescopes was solved for the Sino-German 
$\lambda$6\ cm polarisation survey of the Galactic plane by \citet {Sun07}, where polarisation data on an
absolute zero-level were not available, by adding scaled large-scale components from the  $WMAP$ K-band 
($\lambda$1.3\ cm) polarisation data \citep{Page07}. The $WMAP$ polarisation data are at an absolute zero-level, as required for this purpose.   
This procedure assumes that Faraday rotation of the large-scale emission in the Galactic plane
has a negligible influence on $\lambda$6\ cm polarisation angles ${PA}$s ($PA = \frac{1}{2} $atan$({U/Q})$ ), and 
thus the ratio of $U$ and $Q$ for large scales remains unchanged. This assumption seems to be valid for most
regions of the Galactic plane at $\lambda$6\ cm, except for some emission from the innermost Galaxy.
The Monogem ring is located in the Galactic anti-centre 
direction and well outside of the Galactic plane, so that applying the same correction method for the Effelsberg $\lambda$11\ cm 
polarisation data seems to be justified although the Faraday rotation is about three times higher than at $\lambda$6\ cm.  

We used the $\lambda$21\ cm and the $\lambda$1.3\ cm $WMAP$ absolute polarisation data (nine-year release, \citet{Hinshaw09}) to 
calculate the spectral 
index $\beta$ of the diffuse large-scale $PI$ outside of the area of the thermal filament from the mean $U$ and $Q$ values.  
We found $\beta$ = -3.1 ($T_\mathrm{b}$ $\propto  \nu^{+\beta}$). Between $\lambda$21\ cm and  $\lambda$1.3\ cm, the
mean angle difference for the large-scale emission is about 4\fdg4$\pm5\degr$. With $PA_{\lambda}$ = $RM \times \lambda^{2}$ + $PA_{0}$,
this angle difference corresponds to $RM$ = 1.7$\pm1.9$~rad m$^{-2}$
and implies a negligible correction for the $\lambda$11\ cm polarisation data. It proves that the method we used for 
zero-level correction can be applied to the $\lambda$11\ cm map.  A recent spectral study of polarised
emission observed with $PLANCK$ between 30~GHz and 44~GHz by \citet{Jew} revealed similar spectral-index values in the range $\beta$ = -2.99 to $\beta$ = -3.12, depending on the method that was applied.  

Based on the spectral extrapolation with $\beta$ = -3.1 from the $\lambda$21\ cm or the $\lambda$1.3\ cm data,
we added zero-level offsets of +20.5~mK and +11.2~mK to the originally observed $\lambda$11\ cm $U$ and $Q$ values. 
The effect on the resulting $PI$ emission is shown in Fig.~\ref{11cm+K9}. The morphology 
changed drastically compared to the $PI$ distribution in Fig.~\ref{11cm}, where  the apparent $PI$ emission along
the H$\alpha$ filament is in emission, into a depression (Fig.~\ref{11cm+K9}). This result clearly demonstrates the 
importance of absolute zero-levels for mapping Galactic polarised emission and its analysis. A depression is expected when 
the H$\alpha$ filament causes sufficiently high Faraday rotation on the polarised background, which then adds to the
foreground emission. When the spectral index varies by $\pm$ 0.1, the quoted zero-level offsets for $U$ and $Q$ vary 
by 20$\%$. This does not change the morphology, but $PI$ and $PA$ values are slightly different. We take the influence
of offset variations into account when we discuss the Faraday-screen model in Sect.~4.

\begin{figure}
\centering
\includegraphics[angle=-90, width=0.49\textwidth]{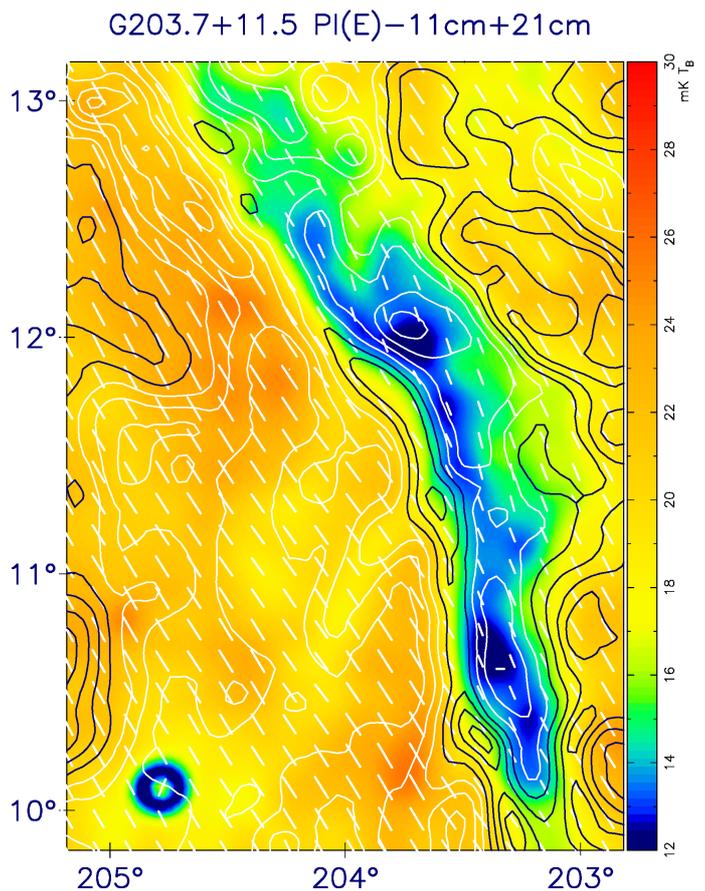}
\caption{Colour-coded $\lambda$11\ cm polarised intensities at 9$\farcm$4 resolution, overlaid with 
contours of $\lambda$21\ cm polarised intensity  starting at 20~mK $T_\mathrm{b}$  in steps of 
20~mK $T_\mathrm{b}$. Up to 80~mK $T_\mathrm{b}$ contours are shown in black.  Intensities of  100~mK $T_\mathrm{b}$ and higher 
are shown by white contours.  Both polarised intensities are at an absolute zero-level.
} 
\label{21cmPI+11cm}
\end{figure}

\section{Total intensities and polarised emission}

The $\lambda$21\ cm and $\lambda$11\ cm total-intensity  contours of Fig.~\ref{21cm} and Fig.~\ref{11cm}
show several compact sources in 
the field and small fluctuations of the diffuse emission. No significant intensity gradient or any 
filamentary structure is visible in total intensities. The polarisation maps at an absolute zero-level (Fig.~\ref{21cm}
and  Fig.~\ref{11cm+K9}), however, 
show  filamentary features that have no counterpart in total intensities and thus result from Faraday-rotation
effects along the line of sight.  The narrow depolarisation canal  as seen in the 
$\lambda$21\ cm polarised-intensity map (Fig.~\ref{21cm}) described above is located at the gradient of the broader 
$\lambda$11\ cm depolarisation filament (Fig.~\ref{21cmPI+11cm})  that extends from {\it{l,b}} $\sim 204\fdg5, +13\degr$ to {\it{l,b}} $\sim 203\fdg1, +10\degr$, which 
itself coincides very well with the faint H$\alpha$ filament (Fig.~\ref{Ha}).

\section{Faraday-screen model for  G203.7+11.5}

\citet{Sun07} have described a method for calculating the properties of a discrete Faraday screen located
somewhere along the line of sight. The modified polarised emission $PI_{\mathrm{on}}$ in the direction of the
Faraday screen (on-position) is compared with the emission outside of the 
screen $PI_{\mathrm{off}}$ (off-position). In addition, the difference of the polarisation angles
$PA_{\mathrm{on}} - PA_{\mathrm{off}}$ is required.
The parameter $c$ is the ratio $PI_{\mathrm{fg}}$/$(PI_{\mathrm{fg}}+PI_{\mathrm{bg}})$, where fg and bg indicate the polarised foreground and 
background components, and $\psi_{s}$ is the angle rotation caused by the Faraday screen. The parameter
$f$ describes the depolarisation of the Faraday screen, where 1 stands for no and 0 for total depolarisation,

\begin{equation}
\centering
\displaystyle{
\left\{
\begin{array}{cc}
\displaystyle
\frac{PI_{\mathrm{on}}}{PI_{\mathrm{off}}}=\sqrt{\mathit{f}^2(1-c)^2+c^2+2\mathit{f}c(1-c)\cos2\psi_s}\ , \\ \displaystyle
PA_{\mathrm{on}} - PA_{\mathrm{off}}=\frac{1}{2}\arctan\left(\frac{\mathit{f}(1-c)\sin2\psi_s}{c+\mathit{f}(1-c)\cos2\psi_s}\right).
 &
\end{array}
\right.
}
\label{eq1}
\end{equation}

For G203.7+11.5, $PA_{\mathrm{off}}$ is not around zero, as is typical of  $PA$s in the Galactic plane  \citep{Sun07}, but at about
$20\degr$, which is an offset to be subtracted from the $PA$ map, so that Eq.~\ref{eq1} can be directly applied. At $\lambda$11\ cm,
we found 
for the two required observables  $PI_{\mathrm{on}}/PI_{\mathrm{off}}$ around 0.5 with variations, where
the depolarisation maximum at {$\it{l,b}$} $\sim 203\fdg7, +11\fdg9$
was used for $PI_{\mathrm{on}}$. We measured for $PA_{\mathrm{on}}-PA_{\mathrm{off}} \sim -15\degr$ to $\sim -20\degr$. 
$PI_{\mathrm{on}}/PI_{\mathrm{off}}$ increases at the gradients of the filament.

There are considerable fluctuations in $PI$ and $PA$. They prevent a very precise estimate of the differences
between on- and off-positions that indicate that the physical properties within the filament vary. In the following, we therefore 
calculate  representative parameters. 
Based on Eq.~\ref{eq1}, we obtained an angle rotation caused by the filament 
in the range of $\psi_{s} \sim -62\degr$ to $ \sim -66\degr$ for $f$ = 1. Then $c$ is calculated as $c$ = 0.63$\pm$0.03,
which is the portion of the foreground polarised emission.
For a wavelength of $\lambda$11\ cm, the corresponding $RM$ is  -86$\pm3$~rad~m$^{-2}$. When the internal depolarisation increases, $c$ and/or parameter $f$ decrease to  $ c$ = 0.61$\pm$0.03 for $f$ = 0.9 and $c$ = 0.58$\pm$0.03 
for $f$ = 0.8, where the corresponding $RM$ changes slightly to -82$\pm3$~rad~m$^{-2}$. 

When we take the spectral uncertainties of $\Delta\beta \pm0.1$ into account when we calculate the $U$ and $Q$ offsets at $\lambda$11\ cm,
we see almost no effect on the angle differences $PA_{\mathrm{on}} - PA_{\mathrm{off}}$, but the 
ratio $PI_{\mathrm{on}}/PI_{\mathrm{off}}$ changes from
about 0.5 for a spectral index of $\beta$ = -3.1 to about 0.6 or 0.4 in the case of a steeper or a flatter spectrum, respectively.
The corresponding foreground portions $c$ do  not change, but the $RM$ values are then calculated as $RM$ =  -96$\pm2$~rad~m$^{-2}$
for  $\beta$ = -3.2 and  $RM$ = -76$\pm2$~rad~m$^{-2}$ for  $\beta$ = -3.0. 

With decreasing depolarisation at the gradients of the filament, the absolute $RM$ values will also decrease and are expected to affect the   
 $\lambda$21\ cm data, where the maximum depolarisation is expected at about $\pm 35$ rad~m$^{-2}$, corresponding to an angle
rotation of $\pm90\degr$ of the polarised background. 
Figure~\ref{21cmPI+11cm} shows that the narrow depolarisation canal runs parallel along the eastern gradient of the 
$\lambda$11\ cm depolarised filament. The $RM$ value of the $\lambda$21\ cm canal of around 
-35~rad~m$^{-2}$ at the outer gradient of the filament is as expected, when the high absolute $RM$ values decrease and match the 
low $RM$ of the diffuse offset-emission. 
Along the $\lambda$11\ cm depolarised filament, the $\lambda$21\ cm $PI$ is at a similar level compared to the off-area, which implies 
that the depolarisation factor $f$ must be close to 1. The depolarisation from the Faraday screen is low and the assumption of 
$f$ = 0.8 we quoted above seems to be a lower limit. 

The slope of the $PI$ gradients of the filament towards longer and lower Galactic longitudes are different, as 
can be clearly seen in the $\lambda$11\ cm $PI$ map at $9\farcm4$ resolution in Fig.~\ref{21cmPI+11cm}. The 
$\lambda$21\ cm canal is visible at the steeper gradient, but on the other side of the filament, there is  a more extended 
$PI$ minimum, as expected from the shallow $\lambda$11\ cm $PI$ and $RM$ gradients, which cannot be clearly
separated from the general $PI$ fluctuations in this area. Thus the maximum depolarisation at around -35 rad~m$^{-2}$ 
is smoothed out. 

\section{Discussion}

\subsection{Thermal filament}

To calculate the physical parameters of G203.7+11.5, we have to know its distance, size, and the electron
temperature of the thermal gas. We can only make estimates of these parameters, which is reflected in the result. 
We assume that G203.7+11.5 is located at the distance of the Monogem ring, which is about 300~pc. The filament extends slightly
to the north and south of the area shown in Fig.~\ref{Ha} as seen in the \citet{Finkbeiner03} H$\alpha$ map. Figure~\ref{Ha} 
corresponds to the area of the $\lambda$11\ cm observations. We estimated a total projected length of the filament of about
$4\fdg5$, which corresponds to 24~pc. However, the apparent size of the Monogem ring is about 25$\degr$, which means a 
diameter of about 130~pc if it is symmetric in 3D. If the G203.7+11.5 filament is not located near the Monogem ring centre, as 
suggested by its coordinates, but instead in the SNR shell, its distance then is about 235~pc or 365~pc. Its projected length in that 
case is about 19~pc or 29~pc, respectively.

The depolarising H$\alpha$ filament has no counterpart in total intensities, which means that the thermal emission
must be very low. From the H$\alpha$ excess, $I_{\mathrm{H\alpha}}$, of 3 to 4~Rayleigh, we may calculate the emission measure 
$EM$[pc~cm$^{-6}$] from Eq.~\ref{eq2}, where $T{_4}$ is the electron temperature in units of 10${^4}$~K.
An $E(B-V)$ extinction correction in the direction of G203.7+11.5 raises  $EM$ by about 2\%  \citep{GGreen19} and can be disregarded 
in view of all other uncertainties,

\begin{equation}
\centering
EM = 2.75~T_{4}^{0.9}~I_{\mathrm{H\alpha}}~exp[2.44E(B-V)].
\label{eq2}
\end{equation}

The range of $EM$ values in the central area of the filament is between 3.6 and 11~pc~cm$^{-6}$ for electron temperatures 
of 4\,000~K and 10\,000~K and H$\alpha$ intensities of 3 to 4~Rayleigh. We adopt an $EM$ of 7 ~pc~cm$^{-6}$ in the following.

The width of the depolarising filament is about 30$\arcmin$, which corresponds to 2.6~pc for a distance of 300~pc. For  
a cylindrical morphology, we obtain the same size $L$ along the line of sight, but if G203.7+11.5 is a sheet-like structure
seen edge-on, 
$L$ may be larger. In the following, we assume the cylindrical case, which means $L$ = 2.6~pc. With
$EM$~[pc~cm$^{-6}$] =  $n_{\mathrm{e}}^{2}$ [cm$^{-6}]\times$ $L$~[pc], 
the average electron density $n_{\mathrm{e}}$ is about 1.6~cm$^{-3}$ for $EM$ = 7~pc~cm$^{-6}$. As discussed above, with the range
of possible distances 
of 235~pc to 365~pc and the range of $EM$ as a function of the electron temperature, we calculated 
$n_{\mathrm{e}}$   between 
1.1~cm$^{-3}$ and 2.3~cm$^{-3}$.  However, for a clumpy thermal gas, the influence
of the filling-factor $f_\mathrm{{n_e}}$ for $n_{\mathrm{e}}$ has to be taken into account. $n_{\mathrm{e}}$ depends on $f_\mathrm{{n_e}}$ as 

\begin{equation}
\centering
n_{e} = \sqrt{\frac{EM}{\mathit{f_\mathrm{{n_{e}}}}L}}~ {\rm cm^{-3}}.
\label{eq5}
\end{equation}

Thus the electron density scales with the filling factor $f_\mathrm{{n_e}}$ as $1 / \sqrt{f_\mathrm{{n_e}}}$. It is unclear what a reasonable filling factor for G203.7+11.5 might be. In any case, values of $f_\mathrm{{n_e}}$ below 1 will increase $n_{\mathrm{e}}$. Recent 
discussions of the filling factor were reported by  \citet{Harvey-Smith11} for \ion{H}{II}-regions and by \citet{Gao15} for the W4 Super-Bubble.

\subsection{Magnetic field strength and pressure of G203.7+11.5}

From the Faraday-screen model, we derived an $RM$ of about $-86$~rad~m$^{-2}$ for the central part of G203.7+11.5. $RM$ depends
on the regular magnetic field strength along the line of sight, the electron density, and the thickness $L$ of the filament
as  $RM$ = 0.81 $n_{\mathrm{e}}$[cm${^{-3}}$] $B{_{||}} [\mu$G] $L$[pc]. We calculated a magnetic field strength along the line of sight
of about 26~$\mu$G with an uncertainty of about 20$\%$. 26~$\mu$G is a lower limit because the orientation of the filament 
is most likely inclined with respect to the line of sight. A filling factor $f_\mathrm{{n_e}}$ that is most likely below 1
will increase $B{_{||}}$ by $B{_{||}} / \sqrt{f_\mathrm{{n_e}}}$. 
This high magnetic field strength largely exceeds that of the local Galactic magnetic field in the
ISM in any case. This local Galactic magnetic field is typically a few $\mu$G,  see for example \citet{Sun08}, \citet{Sun10}, \citet{Ferriere11}, \citet{vanEck11}, 
and \citet{Farrar12}, 
by an order of magnitude, but is not unusual for the magnetic field strength expected in SNR shock fronts. 
Beyond the compression of the ambient Galactic magnetic field in the adiabatic SNR expansion phase by a factor of four, further magnetic field 
amplification effects may increase the magnetic field strength in SNR shock fronts up to 100~$\mu$G or more, see for example
\citet{Reynolds12} and \citet{Dubner15}.

The magnetic field strength in thermal filaments traced by H$\alpha$ emission is not known, but may be compared with filaments found 
for other Faraday screens. The study of magnetic fields of large \ion{H}{II} regions based on $RM$s of extragalactic sources
by \citet{Harvey-Smith11}
revealed values of the regular magnetic field component along the line of sight, $B_{||}$, of between 2 and 6~$\mu$G. This
clearly is lower than
the values we found for G203.7+11.5 and similar to Galactic magnetic field strengths observed in the disc.

Several Faraday screens were detected and discussed in the $\lambda$6\ cm Urumqi survey publications
by  \citet{Sun07}, \citet{Sun11a}, \citet{Gao10}, and \citet{Xiao09}.
High $RM$ values were found for some \ion{H}{II} regions, and in particular, from a few nearly spherical Faraday screens with sizes
of up to 
several degrees. For most of these Faraday screens, the thermal electron density must be very low
because the thermal emission is not visible in H$\alpha,$ 
and moreover, its radio continuum emission is too faint to be detected. Thus, the electron density could not be precisely determined
and just lower limits for $B_{||}$ can be quoted, which reach values up to 10~$\mu$G. 

From a Faraday-screen analysis of the large W4 Super-Bubble, \citet{Gao15} derived $B_{||}$ of 5~$\mu$G and estimated that the 
total magnetic field strength will exceed 12~$\mu$G when its geometry is taken into account.
\citet{Wolleben04} found excessive Faraday rotation towards the ionised rims of some local Taurus molecular clouds
and derived  values for $B_{||}$ 
that exceed 20~$\mu$G. The last two results for $B_{||}$ are close to what we found for G203.7+11.5. 
All Faraday-screen results, with the exception of those for \ion{H}{II} regions, indicate
regions or objects in the Galaxy with a significantly enhanced regular
magnetic field strength when compared to typical Galactic values. In most cases, the origin is not clear, 
although for G203.7+11.5, the old SNR shock-front of 
the Monogem ring seems to be a good candidate for having caused its strong magnetic field.

We calculated the magnetic pressure $P_\mathrm{{mag}} = B_\mathrm{{tot}}^{2}/8\pi$ for G203.7+11.5 as
$P_\mathrm{{mag}}$ =2.7${\pm}0.5{\times}$10$^{-11}$\ dyn\ cm$^{-2}$. 
The thermal pressure $P_\mathrm{{ther}} = 2n_{0}kT_{e}$, where $n_{0} = n_{e}$, in the case of total ionisation,
and $T_{e}$ is taken as 7\,000~K. We determined $P_\mathrm{{ther}}$ = 3.1${\pm}1.2{\times}$10$^{-12}$\ dyn\ cm$^{-2}$.
Clearly, the magnetic pressure largely exceeds the thermal pressure and thus determines the shape and evolution of G203.7+11.5. 

\subsection{$RM$s of extragalactic sources and pulsars compared to the $RM$ of G203.7+11.5}

$RM$s of extragalactic sources in the G203.7+11.5 area were selected from the catalogue by \cite{Xu14}.
In general, the $RM$s in the field are positive. Seven listed $RM$s are in the range +72 to +100~rad/m$^2$. For the 
source at ${\it l,b} = 204\fdg8,10\fdg1$, four $RM$s with +7.0,+7.0,-7.7 and -74.4~rad~m$^{-2}$ were listed, and the source
at ${\it l,b} = 205\fdg1,12\fdg98$ has $RM$ = +3.5 rad~m$^{-2}$. However, these two sources have the largest distances
from the map centre, so that most of the sources probably indicate that the large-scale magnetic field direction
points towards us. The nearby pulsar PSR B0656+14 in the centre of the Monogem ring also has a positive but lower $RM$ of
$RM$ = +22.73 rad m$^{-2}$ \citep{Sobey19}. \citet{Vallee84} have modelled the $RM$ distribution of the Monogem ring.
They calculated a positive $RM$ excess for a thick-shell object when compared with its surroundings. The $RM$ value 
we found above from comparing the diffuse polarised emission at $\lambda$21\ cm and at  $\lambda$1.3\ cm (K band)   
of about $RM$ = 1.7$\pm1.9$~rad m$^{-2}$ is much lower and indicates that very local diffuse synchrotron emission dominates.
Thus, the $RM$s of extragalactic sources in general trace the magnetic field beyond the Monogem ring.    

\subsection{Origin of the magnetic field of G203.7+11.5}

The G203.7+11.5 filament has an $RM$  with a negative sign, and thus its magnetic field points in the opposite direction 
compared to the large-scale Galactic field traced by extragalactic sources. This means that the magnetic field of G203.7+11.5 
does not result from a simple compression of the large-scale field. Shock-excited optical filaments in the Monogem area have been identified 
previously, and G203.7+11.5 may have the same origin, although optical spectra to prove its shock excitation are 
missing so far. The opposite magnetic field direction may result from instabilities that arise from the interaction of the SNR shell 
with interstellar clouds or from the reverse-shock interaction with material from the SN explosion if the filament is located 
close to the centre of the Monogem ring. 
Investigations of Rayleigh-Taylor instabilities by \citet{Jun95} showed that filaments that lie orthogonal to the direction of
the expanding shock-wave may be formed in this way, and magnetic fields will also be amplified by this process. It is important to find signs of shock excitation for G203.7+11.5, which will strongly support the suggested association with the 
Monogem ring from the present study. 

\subsection{Local synchrotron emissivity}

We can use the separation of the foreground and background polarised emission to constrain the local synchrotron emissivity in the 
direction of the Monogem ring. Its small distance allows us to check previous claims of a local excess of synchrotron emissivity 
\citep{Fleishman95, Wolleben04}. The origin of the excess is not understood, in particular if it is caused by an enhancement of 
the local cosmic-ray electron density or by a stronger  magnetic field, when compared to more distant emissivity values. The 3D structure of 
the local excess is not known either, but more data from different directions will provide key information to determine its location.

We calculated the $\lambda$21\ cm synchrotron emissivity up to the 300~pc distance of G203.7+11.5 in the following way: The percentage 
of polarised foreground emission at  $\lambda$11\ cm was determined from the Faraday-screen model to be about 63$\%$ of the total
polarised emission. We assumed the same percentage at  $\lambda$21\ cm, which gives about 70~mK $T_\mathrm{b}$ polarised foreground 
emission. For the maximum possible degree of polarisation of 70$\%$, 
the total intensity is about  100~mK  $T_\mathrm{b}$ or 330~mK $T_\mathrm{b}$/kpc. A more realistic polarisation of 20$\%$ increases 
the intensity to about 350~mK  $T_\mathrm{b}$ or 1.1~K $T_\mathrm{b}$/kpc.  The total-intensity emission along the line of sight at $\lambda$21\ cm 
is about 0.7~K $T_\mathrm{b}$ and the polarisation is about 16$\%$, so that the assumption of 20$\%$ polarised emission for the local 
emission seems realistic. The value of 1.1~K $T_\mathrm{b}$/kpc is below the value of 1.7~K $T_\mathrm{b}$/kpc derived by \citet{Wolleben04} 
for the foreground synchrotron emission based on Faraday-screen effects at the rims of the local Taurus molecular clouds at a distance of 140pc,
but confirms that the local synchrotron emissivity is higher than more distant Galactic values. \citet{Wolleben04} have discussed available 
Galactic synchrotron emissivity data, which span a range from 0.14~K $T_\mathrm{b}$ to 0.9~K $T_\mathrm{b}$ at $\lambda$21\ cm. The recent study 
of low-frequency absorption data from \ion{H}{II}~regions by \citet{Poldermann19} confirmed an increased local synchrotron emissivity.

\section{Summary}
Radio continuum and polarisation observations at  $\lambda$11\ cm and $\lambda$21\ cm were used to study the physical properties 
and the magnetic field of the thermal filament G203.7+11.5.  G203.7+11.5 acts as a Faraday screen and is most likely related 
to the Monogem ring,  which is classified as an evolved SNR.  For a distance of about 300~pc and an 
electron density of 1.6$\pm0.6$~cm$^{-3}$, we found a regular magnetic field component along the line of sight of about  
26$\pm5$~$\mu$G, which is a lower limit for the total regular field in the filament.  
The resulting $RM$ of -86$\pm3$~rad m$^{-2}$ is similar to the $RM$s of extragalactic sources observed in this 
direction, but with an opposite sign. The magnetic field direction of G203.7+11.5 and its unusual strength most likely originate in 
shock interaction of the Monogem ring with clumpy interstellar material, where instabilities cause locally enhanced magnetic fields. 
\\

\begin{acknowledgements}
This research is based on observations with the Effelsberg 100-m telescope of the MPIfR, Bonn.
X.S. is supported by the National Natural Science Foundation of China (Grant No. 11763008).

\end{acknowledgements}

\bibliographystyle{aa}
\bibliography{bbfile}

\end{document}